\documentstyle[12pt]{article}
\parskip=\medskipamount

\def\appendix#1{
\addtocounter{section}{1}
\setcounter{equation}{0}
\renewcommand{\thesection}{\Alph{section}}
\section*{Appendix \thesection\protect\indent #1}
\addcontentsline{toc}{section}{Appendix \thesection\ \ \ #1}
}
\newcommand{\tr}[1]{\,{\rm tr}\,#1\,}

\def\be{\begin{equation}}
\def\la{\label}
\def\ee{\end{equation}}
\def\bea{\begin{eqnarray}}
\def\eea{\end{eqnarray}}
\def\eps{\varepsilon}
\def\a{\alpha}
\def\b{\beta}

\def\n{\nabla}

\def\D{\Delta}
\def\G{\Gamma}
\def\O{\Omega}
\def\g{\gamma}
\def\d{\delta}

\def\l{\left(}
\def\r{\right)}
\def\p{\partial}
\def\x{\vec{x}}

\def\w{\vec{w}}

\def\vk{\vec{k}}

\newcommand{\k}{{\kappa}}

\newcommand{\cL}{{\cal L }}
\newcommand{\La}{\Lambda }

\begin{document}
\begin{titlepage}
\thispagestyle{empty}
\title{
\large{
\begin{flushright}
LMU-TPW 007\\
UAHEP002\\
hep-th/0002170
\end{flushright} }
\vskip 0.5cm
{\bf \large{
Four-point Functions of Lowest Weight CPOs \\ 
in ${\cal N}=4$ SYM$_4$ in Supergravity Approximation
} }}  
\author{G.Arutyunov$^{a\, c}$
\thanks{arut@theorie.physik.uni-muenchen.de} \mbox{}
 and \mbox{} S.Frolov$^{b\,c}$
\thanks{frolov@bama.ua.edu 
\newline
$~~~~~$$^c$On leave of absence from 
Steklov Mathematical Institute, Gubkin str.8, GSP-1, 117966, Moscow, Russia
}
\vspace{0.4cm} \mbox{} \\
\small {$^a$ Sektion Physik,}
\vspace{-0.1cm} \mbox{} \\
\small {Universit\"at M\"unchen,}
\small {Theresienstr. 37,}
\vspace{-0.1cm} \mbox{} \\
\small {D-80333 M\"unchen, Germany}
\vspace{0.4cm} \mbox{} \\
\small {$^b$ Department of Physics and Astronomy,}
\vspace{-0.1cm} \mbox{} \\
\small {University of Alabama, Box 870324,}
\vspace{-0.1cm} \mbox{} \\
\small {Tuscaloosa, Alabama 35487-0324, USA}
\mbox{}
}
\date {}
\maketitle
\begin{abstract}
We show that the recently found quartic action for the scalars 
from the massless graviton multiplet of type IIB supergravity 
compactified on $AdS_5\times S^5$ background coincides 
with the relevant part of the action of the gauged ${\cal N}=8$
5d supergravity on $AdS_5$. We then use this action to compute 
the 4-point function of the lowest weight chiral primary operators 
$\tr(\phi^{(i}\phi^{j)})$ in ${\cal N}=4$ SYM$_4$ at large $N$
and at strong `t Hooft coupling.  
\end{abstract}
\end{titlepage}

\section{Introduction}
The AdS/CFT duality \cite{M,GKP,W} provides a 
remarkable way to approach 
the problem of studying correlation functions in 
certain conformal field 
theories. For ${\cal N}=4$ supersymmetric Yang-Mills 
theory in 
four dimensions (SYM$_4$) this duality allows 
one to find the generating functional 
of Green functions of some composite gauge 
invariant operators at large $N$
and at strong `t Hooft coupling $\lambda$ by 
computing the on-shell value of the 
type IIB supergravity action on $AdS_5\times S^5$ 
background  \cite{GKP,W}.

Thus, the knowledge of type IIB supergravity 
action up to n-th order 
in perturbation of fields near their background 
values is a necessary starting point 
for computing $n$-point correlation functions of 
corresponding operators 
in SYM$_4$. At present the quadratic \cite{AF3} and 
cubic \cite{LMRS,AF5,Lee}
actions for physical fields of type IIB supergravity 
are available that 
allows one to determine normalizations for many 
two- and three-point
functions.

With four-point functions the situation is much more involved \cite{LT1}-\cite{HPR2}. 
So far the only known examples here are the 4-point functions  
of operators $\tr (F^2+\cdots )$ and $\tr (F\tilde{F} +\cdots )$
  \cite{LT1,HFMMR} that on the gravity 
side correspond to massless modes of dilaton and axion fields,
where the relevant part of the gravity action was known. 
These operators are 
rather complicated, in particular, in representation of 
the supersymmetry algebra
they appear as descendents of the primary operators $O_2^{I}=\tr(\phi^{(i}\phi^{j)})$,
where $\phi^i$ are Yang-Mills scalars transforming in the fundamental 
representation of the $R$-symmetry group $SO(6)$. 
The descendent nature of these operators
brings considerable complications both 
in perturbative analysis of the correlation functions, 
and in study of their Operator Product Expansion (OPE) 
from AdS gravity \cite{HMMR}.   

More generally in ${\cal N}=4$ SYM$_4$  there are chiral multiplets 
generated by (single-trace) chiral primary operators (CPO):
$O_k^{I}=\tr (\phi^{(i_1}\cdots\phi^{i_k)})$,  
transforming in the $k$-traceless symmetric representation of $SO(6)$. 
Eight from sixteen supercharges 
annihilate $O^I_k$ while the other eight 
generate, under supersymmetry transformations, the chiral multiplets. 
A fundamental property of CPOs is that they have 
conformal dimensions protected against quantum corrections. 
Thus, they may be viewed as BPS states 
preserving 1/2 of the supersymmetry.
In particular, the lowest component CPOs  
$O_2^{I}$ comprise together 
with their descendents a multiplet containing 
the stress-energy tensor and 
the $R$-symmetry current. 

Recently we have found the quartic effective 5d action 
for scalar fields $s^I$
that correspond at linear order to chiral primary 
operators $O^I$ \cite{AF6}. 
We have also shown that the found action 
admits a consistent 
Kaluza-Klein (KK) truncation to fields from 
the massless graviton multiplet.
This multiplet represents a field content of the 
gauged ${\cal N}=8$, $d=5$ supergravity 
\cite{GRW1,PPN,GRW} and by 
the AdS/CFT correspondence it is dual to the Yang-Mills 
stress-energy multiplet.

Clearly, these results provide a possibility to find 
four-point functions 
of {\it any} CPOs\footnote{The fields $s^I$ 
correspond to extended CPOs
involving single- and multi-trace CPOs and 
their descendents, see
 \cite{AF5,AF6}.
However, for generic values of conformal 
dimensions CPOs and 
extended CPOs have the same correlation functions.} 
in supergravity approximation.
In this paper as the first step in this direction we compute 
the simplest four-point correlation functions 
for all lowest weight CPOs $O_2^{I}$. Hopefully, this will further
extend our understanding of the OPE
in ${\cal N}=4$ SYM$_4$ at strong coupling. 
The detailed study of the OPE of two lowest weight CPOs 
will be the subject 
of a separate paper.   

We start by showing that the quartic action \cite{AF6}
found by compactifying IIB supergravity on the 
$AdS_5\times S^5$ with the further 
reduction to the massless multiplet coincides 
after some additional field redefinitions 
with relevant part of the action for the gauged ${\cal N}=8$
five-dimensional supergravity on $AdS_5$.
This fact together with consistency of the KK reduction 
demonstrates, in particular, that within the supergravity approach, 
four-point correlation functions for fields from 
the YM stress-energy multiplet 
are completely determined by the 5d gauged 
supergravity, i.e., they do not receive 
any contributions from higher KK modes.  

The gauged ${\cal N}=8$ five-dimensional supergravity 
has 42 scalars with 20 of them forming a singlet of the global 
invariance group 
$SL(2,{\bf R})$. These 20 scalars $s^I$ comprise the {\bf 20} 
irrep. of $SO(6)$
and correspond to CPOs $O^I=C^I_{ij}\tr(\phi^i\phi^j)$, 
where $C^I_{ij}$
is a traceless symmetric tensor of $SO(6)$. As we will see the 
only fields that appear in Feynman exchange diagrams describing 
the contribution to the 4-point function of $O^I$ are 
the scalars $s^I$,
the graviton and the massless vector fields. There are 
also contributions 
of contact diagrams corresponding to quartic 
couplings of $s^I$ with two-derivatives
and without derivatives.       

The paper is organized as follows. In Section 2 
we summarize the results of the KK reduction obtained in \cite{AF6}
and put the action in a form suitable for comparison 
with the action of gauged 5d supergravity. In Section 3 
we employ an explicit parametrization 
for the coset space $SL(6,{\bf R})/SO(6)$ to write down the relevant 
part of the action 
for gauged 5d supergravity. We then decompose this 
action near $AdS_5$
background solution and after an additional field redefinition 
find an exact agreement with the action obtained by the KK reduction.
Finally in Section 3 we combine our knowledge of the action 
with the technique \cite{HFMMR} of computing exchange 
Feynman diagrams 
over the AdS space 
and give an answer for the 4-point function of lowest weight 
CPOs in terms of 
universal $D$-functions. Some technical details are 
relegated to two Appendices. 
\section{Results of the reduction}
\setcounter{equation}{0}
As was discussed in the Introduction, the computation of 
a four-point function 
of arbitrary CPOs requires the construction of the 
effective 5d gravity action 
with all cubic terms involving two fields $s^I$ 
and with all $s^I$-dependent quartic terms, 
the problem that has been completely solved in \cite{AF6}. 
For the simplest case 
of lowest weight CPOs the corresponding gravity fields are 20 scalars  
$s^I$ with the lowest AdS-mass $m^2=-4$ and they are in the massless 
graviton multiplet. If we restrict our attention to these fields  $s^I$ 
then the relevant part of the action may be written in the form \cite{AF6}:
\bea
S(s)=\frac{4N^2}{(2\pi )^5}\int d^{5}x\sqrt{-g_a} ~ &\biggl (&
\cL_2(s)+\cL_2(\varphi_{\mu\nu} )+\cL_2(A_{\mu})
\nonumber\\
&+&\cL_3(s)+\cL_3(\varphi_{\mu\nu} )
+\cL_3(A_{\mu})+\cL_4^{(0)}+\cL_4^{(2)}\biggr ) ,
\la{action}
\eea
where $g_a$ denotes the determinant of the AdS metric with the signature
$(-1,1,...,1)$:
$$ds^2 = \frac{1}{z_0^2}(dz_0^2 + \eta_{ij}dx^idx^j).$$

The quadratic actions for the scalars $s^I$, the graviton and the massless vector fields on the AdS
space are given by \cite{AF3}
\bea
\cL_2(s)&=& 
\frac{2^8}{3}\sum_I \left( -\frac12 \n_{\mu}s^I\n^{\mu}s^I
-\frac12 m^2s_I^2\right),
\la{as2}\\
\cL_2(\varphi_{\mu\nu})&=& 
-\frac{1}{4}\n_{\rho}\varphi_{\mu\nu}\n^{\rho}\varphi^{\mu\nu}+
\frac{1}{2}\n_{\mu}\varphi^{\mu\rho}\n^{\nu}\varphi_{\nu \rho}-
\frac{1}{2}\n_{\mu}\varphi_{\rho}^{\rho}\n_{\nu}\varphi^{\mu\nu}
\nonumber\\
&+&\frac{1}{4}\n_{\rho}\varphi_{\mu}^{\mu} \n^{\rho}\varphi^{\nu}_{\nu}
+\frac{1}{2}\varphi_{\mu\nu}\varphi^{\mu\nu}+
\frac{1}{2}(\varphi_{\mu}^{\mu})^2,
\la{agr2}\\
\cL_2(A_{\mu})&=&-\frac{1}{12}\sum_I (F_{\mu\nu}(A^I))^2.
\la{aA2}
\eea
Here the field strength $F_{\mu\nu}(A^I)$
is defined by 
$F_{\mu\nu}(A^I)=\p_{\mu}A_{\nu}^I-\p_{\nu}A_{\mu}^I$, 
where $A_{\mu}^I$ 
with $I=1,...,15$ represent 15 massless vectors that correspond
to the Killing vectors of $S^5$. 
All these fields occur in the bosonic part of 
the massless graviton multiplet 
of compactified type IIB supergravity on $AdS_5\times S^5$.
 
The relevant cubic terms can be easily extracted 
from \cite{LMRS,AF5,Lee}, and they are given by 
\bea
&&\cL_3(s)=\frac{5\cdot 2^{11}}{3^3}a_{I_1I_2I_3}~s^{I_1}s^{I_2}s^{I_3},
\la{sss}
\\
&&\cL_3(\varphi_{\mu\nu})=\frac{2^7}{3\pi^{3/2}}~\biggl ( 
\n^{\mu} s^{I}\n^{\nu} s^{I}\varphi_{\mu\nu} 
-\frac12 \left( \n^{\mu} s^{I}\n_{\mu} s^{I}-4 s^{I}s^{I}\right)\varphi_{\nu}^{\nu}\biggr )
\la{ssgr}
\\
&&\cL_3(A_{\mu})=\frac{2^8}{3^2}t_{I_1I_2I_3}~
s^{I_1}\n^{\mu} s^{I_2}A_{\mu}^{I_3}.
\la{ssA}
\eea
Here the summation over $I_1,~I_2,~I_3$ running over 
the basis of irrep. {\bf 20} of $SO(6)$ is assumed, 
and we use the following notations
\bea
a_{I_1I_2I_3}=\int~Y^{I_1}Y^{I_2}Y^{I_3},\quad
t_{I_1I_2I_3}=\int~\n^\a Y^{I_1}Y^{I_2}Y_\a^{I_3},
\nonumber
\eea
where the scalar $Y^I$ and the vector $Y_\a^I$ spherical 
harmonics
\footnote{In this Section
$\a$ is used to denote the index of $S^5$.} 
of $S^5$ 
satisfy $\n_{\a}^2 Y^I=-12 Y^I$, $(\n_{\g}^2-4)Y_\a^I=-8Y_\a^I$.
We also assumed that the spherical harmonics of different types 
are orthonormal, i.e. $\int Y^IY^J=\d^{IJ}$ 
and $\int Y^I_{\a}Y^J_{\a}=\d^{IJ}$.

Finally, in \cite{AF6} the following values of the  quartic couplings  
of the 2-derivative vertex
\bea
\cL^{(2)}_{4}&=&\frac{5^2\cdot 2^9}{27}
\sum_{I_5}a_{I_1I_2I_5}a_{I_3I_4I_5}\n_{\mu}
(s^{I_1}s^{I_2})\n^{\mu}(s^{I_3}s^{I_4})
+\frac{2^{13}}{27\pi^3}\n_{\mu}(s^{I_1}s^{I_1})\n^{\mu}(s^{I_2}s^{I_2})
\eea
and of the non-derivative vertex
\bea
\cL^{(0)}_{4}=-\frac{5^2\cdot 2^{11}}{9}
\sum_{I_5}a_{I_1I_2I_5}a_{I_3I_4I_5}s^{I_1}s^{I_2}s^{I_3}s^{I_4}
\eea 
were found. 

The quartic action can be further simplified by 
substituting the integrals of spherical harmonics
for their explicit value via $C$-tensors (see Appendix A). 
Indeed, by using (\ref{int}) together with 
summation formula (\ref{sum1}) one gets 
\bea
\la{sumaa}
\sum_{I_5}a_{I_1I_2I_5}a_{I_3I_4I_5}
=\frac{2^4\cdot 3}{5^2\pi^3}\l C^{I_1I_2I_3I_4}+C^{I_1I_2I_4I_3}
-\frac{1}{3}\d^{I_1I_2}\d^{I_3I_4}\r . 
\eea
where the shorthand notation 
$C^{I_1I_2I_3I_4}=C^{I_1}_{i_1i_2}C^{I_2}_{i_2i_3}
C^{I_3}_{i_3i_4}C^{I_4}_{i_4i_1}$
for the trace product of four matrices $C^{I}$ was introduced.

By using this formula, the two-derivative Lagrangian may be reduced
to the following form:
\bea
\cL^{(2)}_{4}&=&\frac{2^{14}}{3^2\pi^3}
C_{I_1I_2I_3I_4}\n_{\mu}(s^{I_1}s^{I_2})\n^{\mu}(s^{I_3}s^{I_4}).
\eea 

From the cubic couplings one can see 
that except the self-interaction, the 
scalars from the massless multiplet interact only via exchange 
by the massless graviton 
$\varphi_{\mu\nu}$ and by the massless vector fields $A_{\mu}^I$. 
Introduce a concise notation 
\bea
S(s)=\frac{N^2}{8\pi^2}\int d^{5}x~\sqrt{-g_a}\cL_{red},
\eea
where the subscript in $\cL_{red}$ stands to remind that action 
$S$ is obtained by dimensional reduction, and we have emphasized the 
5-dimensional gravitational coupling $2\k_5^2=\frac{8\pi^2}{N^2}$.

Substituting in (\ref{sss})-(\ref{ssA})
explicit values (\ref{int}) of $a_{I_1I_2I_3}$ and $t_{I_1I_2I_3}$,
using for  $\cL^{(0)}_{4}$ summation formula (\ref{sumaa}), and
rescaling the fields as 
$$
s^I \to \frac{3^{1/2}\pi^{3/2}}{2^{9/2}}s^I ,~~~
A_{\mu}^I\to 6^{1/2}\pi^{3/2}A_{\mu}^I,~~~ 
\varphi_{\mu\nu}\to \pi^{3/2}\varphi_{\mu\nu} ,
$$ we get the Lagrangian  
\bea
\la{ac4p}
\cL_{red}&=&-\frac{1}{4}\l \n_{\mu}s^I\n^{\mu}s^I-4 s^I s^I \r 
+\frac{1}{3}C_{I_1I_2I_3}s^{I_1}
s^{I_2}s^{I_3}\\
\nonumber
&+&\frac{1}{4}\l \n^{\mu}s^I \n^{\nu} s^I \varphi_{\mu\nu}
-\frac{1}{2}\l \n^{\mu}s^I \n_{\mu} s^I-4s^Is^I \r \varphi_{\nu}^{\nu} \r \\
\nonumber
&+&\frac{1}{2^4}C_{I_1I_2I_3I_4}\n_{\mu}(s^{I_1}s^{I_2})
\n^{\mu}(s^{I_3}s^{I_4}) 
-\frac{3}{2^2}C_{I_1I_2I_3I_4} s^{I_1}s^{I_2}s^{I_3}s^{I_4} 
+\frac{1}{2^3}s^{I_1}s^{I_1}s^{I_2}s^{I_2}\\
\nonumber
&+&T_{I_1I_2I_3}s^{I_1}
\n^{\mu}s^{I_2}A_{\mu}^{I_3}
-\frac{1}{2}F_{\mu\nu}^IF^{\mu\nu~I}+\cL_2(\varphi_{\mu\nu}) 
\eea
that will be used in Section 4 to compute the 4-point functions
of the lowest weight CPOs.

Finally we put this Lagrangian in the form most suitable for comparison  
with the relevant part of the action of the 
gauged ${\cal N}=8$ 5d supergravity. 
Introducing the matrices 
\bea
\nonumber
\La=(\La)_{ij}=C_{ij}^Is^I,~~~A_{\mu}=(A_{\mu})_{ij}=-C_{i;j}^IA_{\mu}^I,
\eea
where $C_{ij}^I$ and $C_{i;j}^I$ are described in the Appendix A, 
one obtains
\bea
\la{redact}
\cL_{red}&=&-\frac{1}{4}\tr{\l \n_{\mu}\La 
\n^{\mu}\La-4\La^2\r}
+\frac{1}{3}\tr{\La^3} \\
\nonumber
&+&\frac{1}{4}\l \tr{\n_{\mu}\La \n_{\nu}\La} 
-\frac{1}{2}g_{\mu\nu}\tr{\l \n_{\g}\La \n^{\g}\La-4\La^2\r}\r \varphi^{\mu\nu}  \\
\nonumber
&+&\frac{1}{2^4} \tr{\l\n_{\mu}\La^2 \n^{\mu}\La^2\r} 
-\frac{3}{2^2} \tr{\La^4}+ \frac{1}{2^3} \l\tr{\La^2}\r^2
\\
\nonumber
&+&\frac{1}{2}\tr{F_{\mu\nu}F^{\mu\nu}}-2\tr{\l\n^{\mu}\La \La A_{\mu} \r}
+\cL_2 (\varphi_{\mu\nu}),
\eea
where $\tr{ F_{\mu\nu}F^{\mu\nu}}=-F_{\mu\nu}^{ij}F^{\mu\nu~ij}$ and normalization 
condition (\ref{norm}) was used.

\section{Lagrangian of gauged 5d supergravity}
\setcounter{equation}{0}
Gauged ${\cal N}=8$ five-dimensional supergravity was constructed in \cite{GRW1,PPN}
by gauging abelian vector fields of the ${\cal N}=8$ Poincar\'e
supergravity. The gauged theory has a local non-abelian $SO(6)$ symmetry,
a local composite $USp(8)$ symmetry and a global $SL(2,{\bf R})$ symmetry.   
The bosonic field content is given by graviton,
fifteen real vector fields $A_{\mu ~ij}$, $i,j=1,\ldots 6$ transforming 
in the adjoint representation of $SO(6)$, 12 antisymmetric tensors
of the second rank and by 42 scalars that in the ungauged theory 
parametrize the non-compact manifold $E_{6(6)}/USp(8)$. 
In what follows we adopt the conventions of \cite{GRW}.

Let $A,B,...=1,\ldots 8$ be the indices of the 
representation ${\bf 27}$ of $E_{6(6)}$
and $a,b,...$ be $USp(8)$ indices that are raised 
and lowered with the symplectic 
metric $\O_{ab}$. Explicitly, an element of 
$E_{6(6)}/USp(8)$ can be described by 
the scalar vielbein $V_{AB}^{~~ab}$ which is $27\times 27$.  
In the gauged theory minimal couplings of the connection $A_{\mu ~ij}$ responsible 
for the local $SO(6)$ symmetry are introduced to all the fields transforming 
linearly under $SO(6)$. The transformation properties of the fields under 
$SO(6)$ are then uniquely specified by the embedding of $SO(6)$ into 
the group $SL(6,{\bf R})$, the latter being a subgroup of $E_{6(6)}$.
Recall that under the subgroup $SL(6,{\bf R})\times SL(2,{\bf R})$
the representation ${\bf 27}$ of $E_{6(6)}$ is decomposed as $27=(15,1)+(6,2)$.  
The components of the vielbein are then denoted as $V_{ij}^{~~ab}$
and $V_{i\a}^{~~ab}$, where  $i,j=1,\ldots 6$ 
are $SL(6,{\bf R})$  
and $\a=1,2$ are $SL(2,{\bf R})$ indices. 

The relevant bosonic part\footnote{We put 
all antisymmetric fields to zero, and changed the overall normalization of 
the Lagrangian in comparison to \cite{GRW}.} 
of the Lagrangian of the gauged 5d 
gravity is of the form
\bea
\la{ags}
\cL=R-\frac{1}{6}P_{\mu~abcd}P_{\mu}^{~abcd}-P
-\frac{1}{2}F_{\mu\nu;ij}F^{\mu\nu;ij}.
\eea
Here $F_{\mu\nu;ij}$ is a $SO(6)$-covariant 
Yang-Mills field strength, 
$P$ is a scalar potential and the tensor 
$P_{\mu abcd}$ is given by 
\bea
\nonumber
P_{\mu~ab}^{~~~~cd}=(V^{-1})^{cd~AB}\n_{\mu}V_{AB~ab}+2Q_{\mu [a}^{[c}\d_{b]}^{d]}
-2g(V^{-1})^{cd~ij}A_{\mu i}^{~~k}V_{kj~ab}-g (V^{-1})^{cd~i\a}A_{\mu i}^{~~j}V_{j\a~ab}
\eea
and it represents a coset element
in the decomposition of the $E_{6(6)}$ Lie 
algebra into an $USp(8)$ 
and a coset part.
In particular, matrix $Q_{\mu [a}^{~~~[c}\d_{b]}^{d]}=
\sum_{k=1}^{36} B_{\mu}^k (T^k)_{ab}^{~cd}$
is an $USp(8)$-connection responsible for the 
local $USp(8)$ symmetry.
Recall that $USp(8)$-connection $B_{\mu}^k$ 
is non-dynamical since 
it does not have a kinetic term.  
Therefore, it can be excluded by using 
its equation of motion 
as in fact is done below.
The dimension of $USp(8)$ is 36 and $T^k$
is a basis of the ${\bf 27}$ irrep. of 
the $USp(8)$ Lie algebra, 
$g$ is the Yang-Mills coupling constant.

Eq.(\ref{ags}) is our starting point to find the action 
for scalars $s^I$
on the $AdS_5$ background. Since the potential for $s^I$ 
was already found
in studying the critical points the only missing piece 
is an explicit 
construction of the kinetic term.
  
To build the kinetic term we need an explicit parametrization 
of the scalar vielbein in terms of $20$ scalar fields that are neutral 
under $SL(2,{\bf R})$. We then employ the parametrization of \cite{GRW}, 
in which 42 scalars are represented by two 
real symmetric traceless matrices $\Lambda_i^{~j}$ 
and $\Lambda_{\a}^{~\b}$, $\a,\b=1,2$
and by a real completely antisymmetric in $i,~j,~k$ 
tensor $\phi_{ijk\a}$
obeying the self-duality condition 
$$
\phi_{ijk\a}=\frac{1}{6}\eps_{\a\b}\eps_{ijklmn}\phi_{lmn\b}.
$$   
Since only $\Lambda$ is a singlet under $SL(2,{\bf R})$ in what 
follows we put $\Lambda_{\a}^{~\b}$ and $\phi_{ijk\a}$ to zero.
Turning off these fields is 
allowed in our specific problem of constructing the action for $s^I$
because the existence of the cubic terms 
containing two $SL(2,{\bf R})$-singlets $s^I$  
and one doublet field is forbidden by the $SL(2,{\bf R})$ symmetry.
In fact, from the point of view of the dimensional 
reduction of type IIB supergravity
matrix $\Lambda_{\a}^{~\b}$ describes zero modes of axion 
and dilaton fields 
while $\phi_{ijk\a}$ encodes the scalars arising 
from the reduction 
of the antisymmetric tensor fields. 

With this parametrization at hand we get 
the following expression for the vielbein $V_{AB}^{ab}$ in the 
$SL(6,{\bf R})\times SL(2,{\bf R})$ basis:
\bea
V^{ij~ab}&=&\frac{1}{4}(\G_{kl})^{ab}S_k^iS_l^j,
\quad
(V^{-1})_{ij~cd}=\frac{1}{4}(\G_{kl})_{cd}(S^{-1})_i^k(S^{-1})_j^l,
\nonumber \\
V_{i\a}^{~ab}&=&\frac{1}{2^{3/2}}(\G_{k\a})^{ab}S_i^k,
\quad
(V^{-1})_{cd}^{~i\a}=-\frac{1}{2^{3/2}}(\G_{k\a})_{cd}(S^{-1})_k^i,
\la{viel}
\eea
where $\G$ are $SO(6)$ $\G$-matrices (see Appendix A) and $S$ is given by $S=e^{\Lambda}$ with
$\Lambda$ being the traceless symmetric $6\times 6$-matrix 
comprising $20$ scalars. 

It is convenient to introduce a matrix $R_{\mu}$:
\bea
\la{R}
R_{\mu}=\n_{\mu}SS^{-1}+g S A_{\mu}S^{-1}.
\eea
Since $\Lambda$ is traceless and $A_{\mu i}^{~j}$ is antisymmetric
this matrix appears to be traceless: $R_{\mu i}^{~~i}=0$.

The scalar kinetic part of Lagrangian (\ref{ags}) 
in parametrization (\ref{viel})
is then computed in the Appendix A and the result looks as follows
\bea
\nonumber
P_{\mu abcd}P_{\mu}^{abcd}&=&\frac{3}{2} \tr{\l R_{\mu}+ R_{\mu}^t \r^2 }.
\eea

Substituting the potential found in \cite{GRW}, we get 
the final answer for the Lagrangian
(for simplicity we omit for the moment the gravity and the gauge terms):
\bea
\la{lS}
\cL=-\frac{1}{4}\tr{\l R_{\mu}+ R_{\mu}^t \r^2 }
+\frac{g^2}{8}\l (\tr{SS})^2-2\tr{(SSSS)} \r .
\eea
Scalar fields $\La_i^{~j}$ transform in the ${\bf 20}$ of $SO(6)$.
We are interested in the maximally supersymmetric vacuum
with only non-trivial bosonic fields,
which implies that the background solution 
is invariant under $SO(6)$.
Thus, at the $SO(6)$ invariant critical point $P_0$ of the potential
the scalar fields should acquire some expectation values 
that are invariant under $SO(6)$. Clearly, the only possibility 
for that is to take $\La_i^{~j}=0$,
i.e., to put $S$ to be the unit matrix. The value of the potential 
is then $P_0=-\frac{3}{4}g^2$ that leads to the equation of motion:
$$
R_{\mu\nu}=\frac{4}{3}P_0=-g^2 g_{\mu\nu}.
$$ 
Thus, the background solution is the anti-de Sitter 
space with the cosmological constant $\lambda=-\frac32 g^2$ and with  
vanishing scalars $\La_i^{~j}$. Decomposition 
of Lagrangian (\ref{lS}) near this background is then easily 
obtained by decomposing $S=e^{\Lambda}$ around $\Lambda=0$.

We find up to the cubic order
\bea
\nonumber
\n_{\mu}SS^{-1}&=&\n_{\mu}\La - \frac{1}{2}(\n_{\mu}\La \La- \La \n_{\mu}\La )
-\frac{1}{2}\La \n_{\mu}\La \La +\frac{1}{6}\n_{\mu}\La^3 , \\
\nonumber
(\n_{\mu}SS^{-1})^t&=&\n_{\mu}\La + \frac{1}{2}(\n_{\mu}\La \La- \La \n_{\mu}\La )
-\frac{1}{2}\La \n_{\mu}\La \La +\frac{1}{6}\n_{\mu}\La^3 .
\eea
By using these formulae, one then gets
\bea
 R_{\mu}+ R_{\mu}^t=2\n_{\mu}\La-\La\n_{\mu}\La\La+\frac{1}{3}\n_{\mu}\La^3+2g[\La,A_{\mu}].
\eea
The terms quadratic in $\La$ cancelled and, therefore, the action 
does not contain cubic in $\La$ terms with two derivatives.

Analogously, for the potential we find
\bea
\nonumber
\frac{g^2}{8}\l (\tr{SS})^2-2\tr{(SSSS)} \r =
g^2\l 3+\tr{\La^2}-\frac{2}{3}\tr{\La^3}
-\frac{5}{3}\tr{\La^4}+\frac{1}{2}(\tr{\La^2})^2 \r .
\eea

To compare action (\ref{ags}) with the one from the previous Section
we have to fix the coupling constant $g$.
It is fixed to be $g^2=4$ by the requirement to  
have the vacuum solution defined by the equation  
$R_{\mu\nu}=-4 g_{\mu\nu}$. Namely this background solution 
 was used to obtain 
the action (\ref{redact}) by compactifying ten-dimensional
type IIB supergravity.
 
Thus, for Eq.(\ref{lS}) up to the fourth order in $\La$ we get
\bea
\nonumber
\cL&=&12-\tr{\l \n_{\mu}\La\n^{\mu}\La-4\La^2 \r }
-\frac{2}{3}\tr{(\n_{\mu}\La\La\La\n^{\mu}\La
-\La\n_{\mu}\La \La\n^{\mu}\La  )} \\
\nonumber 
&-&\frac{8}{3}\tr{\La^3}
-\frac{20}{3}\tr{\La^4}+2(\tr{\La^2})^2-
8\tr{\l \n^{\mu}\La\La A_{\mu}\r } 
\eea
It is then useful to perform the following field redefinition 
\bea
\nonumber
\La\to \La+r\La^3
\eea
under which the Lagrangian transforms into 
\bea
\nonumber
\cL&=&R+12-
\tr{\l \n_{\mu}\La\n^{\mu}\La-4\La^2 \r }\\
\nonumber
&-&4\tr{\l \l \frac{1}{6}+r \r \n_{\mu}\La\La\La\n^{\mu}\La
+\l \frac{r}{2}-\frac{1}{6} \r \La\n_{\mu}\La \La\n^{\mu}\La  \r } \\
\nonumber 
&-&\frac{8}{3}\tr{\La^3}
+4\l -\frac{5}{3}+2r\r \tr{\La^4}+2(\tr{\La^2})^2 \\
\nonumber
&+&\frac{1}{2}\tr{F_{\mu\nu}F^{\mu\nu}}
-8\tr{\l \n^{\mu}\La\La A_{\mu}\r },
\eea
where we have restored the gravity and gauge terms.
Let us choose $r$ to be $r=-2/3$. Then taking into account that 
$$
\tr{\l\n_{\mu}\La^2 \n^{\mu}\La^2\r}=2
\tr{\l \n_{\mu}\La\La\La\n_{\mu}\La + \La\n_{\mu}\La \La\n_{\mu}\La  \r },
$$ 
and making the rescaling $\La\to -\frac{1}{2}\La $, we find 
\bea
\la{sugraact}
\cL&=&R+12-\frac{1}{4}\tr{\l \n_{\mu}\La\n^{\mu}\La-4\La^2 \r }
+\frac{1}{3}\tr{\La^3}\\
\nonumber
&+&\frac{1}{2^4}\tr{\l\n_{\mu}\La^2 \n^{\mu}\La^2\r}   
-\frac{3}{2^2} \tr{\La^4}+\frac{1}{2^3} \l\tr{\La^2}\r^2\\
\nonumber
&+&\frac{1}{2}\tr{F_{\mu\nu}F^{\mu\nu}}
-2\tr{\l \n^{\mu}\La\La A_{\mu}\r }.
\eea   
Note that $-6$ is the cosmological constant in the action 
$\int d^{d+1}x\sqrt{-g}(R-2\lambda)$,
$\lambda=-\frac{1}{2}d(d-1)$ for $d=4$ that 
appears in the reduction from ten dimensions. 

Multiplying (\ref{sugraact}) by $\sqrt{-g}$, and decomposing 
the metric $g_{\mu\nu}=g_{\mu\nu}^0+\varphi_{\mu\nu}$
near the background AdS solution $g_{\mu\nu}^0$, one  
immediately finds 
$$
\cL =\cL_{red}.
$$
Thus, we have shown that the action for the scalars $s^I$ obtained by 
compactification of type IIB supergravity on $AdS_5\times S^5$ 
with further reduction to the fields from the massless graviton multiplet 
coincides with the relevant part of the action of the gauged ${\cal N}=8$
five-dimensional supergravity on $AdS_5$ background.

\section{4-point function of lowest weight CPOs}
\setcounter{equation}{0}
The normalized lowest weight CPOs in ${\cal N}=4$
SYM$_4$ are operators of the form 
$$
O^I(\x)=\frac{2^{3/2}\pi^2 }{\lambda}C_{ij}^I\tr(:\phi^i\phi^j:).
$$
By using the following propagator 
$\langle \phi_a^i\phi^j_b \rangle = 
\frac{g_{YM}^2\d_{ab}\d^{ij}}{(2\pi)^2x_{12}^2}$,
where $a,b$ are color indices and $x_{ij}=\x_i-\x_j$, one finds in the free 
approximation and at leading order in $1/N$ the following 
expressions for 2-, 3- \cite{LMRS} and 4-point functions of $O^I$:
\bea
&&\langle O^{I_1}(\x_1)O^{I_2}(\x_2)  \rangle =\frac{\d^{I_1I_2}}{x_{12}^2},
\nonumber  \\
&&\langle O^{I_1}(\x_1)O^{I_2}(\x_2)O^{I_3}(\x_3) 
\rangle =\frac{1}{N}\frac{2^{3/2}C^{I_1I_2I_3}}{x_{12}^2x_{13}^2 x_{23}^2}, \\
\nonumber  \\
\nonumber
&&\langle O^{I_1}(\x_1)O^{I_2}(\x_2)O^{I_3}(\x_3) O^{I_4}(\x_4)\rangle
=\frac{\d^{I_1I_2}\d^{I_3I_4} }{x_{12}^4x_{34}^4}+
\frac{1}{N^2}\frac{4C_{I_1I_2I_3I_4}}{x_{12}^2 x_{14}^2 x_{23}^2 x_{34}^2 }
+\mbox{permutations},
\eea
where the first term in the 4-point function represents the contribution 
of disconnected diagrams.

In this Section we compute 4-point functions of $O^I$ from 
AdS supergravity. The starting point is action (\ref{ac4p}). 
We will work with the Euclidean version of $AdS_5$
that amounts to changing in (\ref{ac4p}) an overall sign,
so that 
\bea
\la{ac4p1}
\cL_{red}&=&\frac{1}{4}\l \n_{\mu}s^I\n^{\mu}s^I-4 s^I s^I \r 
-\frac{1}{3}C_{I_1I_2I_3}s^{I_1}
s^{I_2}s^{I_3}\\
\nonumber
&-&\frac{1}{4}\l \n^{\mu}s^I \n^{\nu} s^I \varphi_{\mu\nu}
-\frac{1}{2}\l \n^{\mu}s^I \n_{\mu} s^I-4s^Is^I \r \varphi_{\nu}^{\nu} \r \\
\nonumber
&-&\frac{1}{2^4}C_{I_1I_2I_3I_4}\n_{\mu}(s^{I_1}s^{I_2})
\n^{\mu}(s^{I_3}s^{I_4}) 
+\frac{3}{2^2}C_{I_1I_2I_3I_4} s^{I_1}s^{I_2}s^{I_3}s^{I_4} 
-\frac{1}{2^3}s^{I_1}s^{I_1}s^{I_2}s^{I_2}\\
\nonumber
&-&T_{I_1I_2I_3}s^{I_1}
\n^{\mu}s^{I_2}A_{\mu}^{I_3}
+\frac{1}{2}F_{\mu\nu}^IF^{\mu\nu~I}-\cL_2(\varphi_{\mu\nu}) 
\eea
It is convenient to introduce the following currents 
\bea
\nonumber
T_{\mu\nu}&=& \n_{\mu}s^I \n_{\nu} s^I 
-\frac{1}{2}g_{\mu\nu}\l \n^{\rho}s^I \n_{\rho} s^I-4s^Is^I \r , \\
\nonumber
J_{\mu}^{I_3}&=&T_{I_1I_2I_3}(s^{I_1}\n_{\mu}s^{I_2}-s^{I_2}\n_{\mu}s^{I_1}),
\eea 
both of them are conserved on-shell: 
$\n^{\mu} T_{\mu\nu}=\n^{\mu}J_{\mu}^I=0$.

From (\ref{ac4p1}) we get the following equations of motion:

\noindent {\bf 1}. for scalars $s^I$:
\bea
\la{eqs}
(\n_{\mu}^2-m^2)s^{I}=-2C_{IJK}s^{J}s^{K};
\eea
\noindent {\bf 2}. for vector fields $A_{\mu}^I$:
\bea
\la{eqv}
\n^{\nu}(\n_{\nu} A_{\mu}^{I}-\n_{\mu} A_{\nu}^{I})=-\frac{1}{4}J_{\mu}^I;
\eea
\noindent {\bf 3}. for the graviton $\varphi_{\mu\nu}$: 
\bea
\la{grst}
W_{\mu\nu}^{~\rho\lambda}\varphi_{\rho\lambda}
=\frac{1}{4}\l g_{\mu\mu'}g_{\nu\nu'}+g_{\mu\nu'}g_{\nu\mu'}
-\frac{2}{3}g_{\mu\nu}g_{\mu'\nu'}\r T^{\mu'\nu'},
\eea
where $W_{\mu\nu}^{~\rho\lambda}$ is the Ricci operator
\bea
\nonumber
W_{\mu\nu}^{~\rho\lambda}\varphi_{\rho\lambda}
=
-\n_{\rho}^2\varphi_{\mu\nu}+\n_{\mu}\n^{\rho}\varphi_{\rho\nu}+\n_{\nu}\n^{\rho}\varphi_{\rho\mu}
-\n_{\mu}\n_{\nu} \varphi_{\rho}^{\rho} 
-2(\varphi_{\mu\nu} - g_{\mu\nu}\varphi_{\rho}^{\rho}).
\eea 
Introduce the scalar $G$ \cite{F}, the vector $G_{\mu\nu}$ and the graviton 
$G_{\mu\nu~\rho\lambda}$ \cite{HFMMR0} propagators
\bea
\nonumber
&&(\n_a^2-m^2)G(u)=-\d(z,w), \\
\nonumber
&&\n^{\rho}(\n_{\rho} G_{\mu\nu}^{I}-\n_{\mu}G_{\nu\rho}^{I})=-g_{\mu\nu}\d(z,w), \\
\nonumber
&&W_{\mu\nu}^{~\rho\lambda}G_{\rho\lambda~\mu'\nu'}=
\l g_{\mu\mu'}g_{\nu\nu'}+g_{\mu\nu'}g_{\nu\mu'}
-\frac{2}{3}g_{\mu\nu}g_{\mu'\nu'} \r \d(z,w)
\eea
being the functions of the invariant AdS-distance $u$:
$$u=\frac{(z-w)^2}{2z_0w_0},\qquad (z-w)^2=\d_{\mu\nu}(z-w)_\mu(z-w)_\nu .$$
Represent the solution to the equations of motion in the form  
$$
s=s^0+s^1,~~~A_{\mu}=A_{\mu}^0+A_{\mu}^1,
~~~\varphi_{\mu\nu}=\varphi_{\mu\nu}^0+\varphi_{\mu\nu}^1,
$$
where $s^0$, $A_{\mu}^0$ and $\varphi_{\mu\nu}^0$ are solutions of the 
linearized equations 
with fixed boundary conditions and  $s^1$, 
$A_{\mu}^1$ and $\varphi_{\mu\nu}^1$ 
are the corrections with vanishing boundary conditions.
Then by perturbation theory for $s^1$, $A_{\mu}^1$
and $\varphi_{\mu\nu}^1$ one gets
\bea
\nonumber
&&
s_{I}^1(w)=2C_{IJK}
\int \frac{d^5z}{z_0^5}G(u)s^{J}(z)s^{K}(z),\\
\la{eqsol}
&&A_{\mu}^{1~I}(w)=
\frac{1}{4} \int \frac{d^5z}{z_0^5}G_{\mu}{}^\nu (u)J^{I}_\nu (z), \\
\nonumber
&&\varphi_{\mu\nu}^1(w)=\frac{1}{4}\int \frac{d^5z}{z_0^5} 
G_{\mu\nu~\mu'\nu'}(u)T^{\mu'\nu'}(z),
\eea
where the r.h.s. depends only on $s^0$, $A_{\mu}^0$
and $\varphi_{\mu}^0$ 
and from now on we omit the superscript $0$ unless 
we want to indicate explicitly 
that we deal with solutions of the linearized equations of motion.

It is worth noting that not only the interaction 
terms but also the 
quadratic action $\cL_{quad}$ gives a contribution 
to the on-shell value of action (\ref{ac4p}) 
depending quartically on $s_0$:
\bea
\nonumber
\cL_{quad}=\frac{1}{2}C_{IJK}
s_0^{I}s_0^{J}s_1^{K}
+\frac{1}{8}\varphi_{\mu\nu}^1T^{\mu\nu}
+\frac{1}{4}A_{\mu}^{1~I}J^{\mu~I}.
\eea

Taking into account the summation formula 
\bea
\sum_{I_5} 
T_{I_1I_2I_5}T_{I_3I_4I_5}=2\l C_{I_1I_2I_4I_3}-C_{I_1I_2I_3I_4}\r ,
\eea
that follows from (\ref{sum2}) and using (\ref{eqs}) we arrive at 
the following expression 
for the on-shell value of (\ref{ac4p1}):
\bea
\nonumber
\cL_{red} &=&\frac{1}{4}C_{I_1I_2I_3I_4}
\int \frac{d^5z}{z_0^5}s^{I_1}
\stackrel{\leftrightarrow~}{\n^{\mu}}s^{I_2}(w)
G_{\mu\nu}(u)s^{I_3} 
\stackrel{\leftrightarrow~}{\n^{\nu}}s^{I_4}(z)\\
\nonumber
&-&\frac{1}{2^5} \int \frac{d^5z}{z_0^5} T^{\mu\nu}(w)
G_{\mu\nu~\rho\lambda}(u)T^{\rho\lambda}(z)
\\
\nonumber
&-&
\l C_{I_1I_2I_3I_4}-\frac{1}{6}\d_{I_1I_2}\d_{I_3I_4}\r
\int \frac{d^5z}{z_0^5}G(u)s^{I_1}(w)s^{I_2}(w)s^{I_3}(z)s^{I_4}(z)
\\
\nonumber
&-&\frac{1}{2^4}
C_{I_1I_2I_3I_4}\n_{\mu}(s^{I_1}s^{I_2})\n^{\mu}(s^{I_3}s^{I_4}) 
+\frac{3}{4}C_{I_1I_2I_3I_4} s^{I_1}s^{I_2}s^{I_3}s^{I_4} 
-\frac{1}{8}s^{I_1}s^{I_1}s^{I_2}s^{I_2} .
\eea
On the language of the Feynman diagrams the first three 
terms here involving 
$z$-integrals describe the exchange by the gauge boson, 
by the graviton and by the scalar
fields respectively. The other contributions 
correspond to contact diagrams.
$z$-integrals are easily computed by the technique of 
\cite{HFR} and in the Appendix B
we list the corresponding results. It is worthwhile to note that 
since we compute the on-shell value of the gravity action, we 
take into account only the connected $AdS$ graphs.

Recall that the solution of the Dirichlet boundary problem 
for the scalar field $s^I$ of 
mass $m^2=-4$ on $AdS_5$ reads as 
\bea
\la{sDp}
s^{I}(z,\x)=\frac{1}{2\pi^2}\int d^4\x K_2(w,\x)s^{I}(\x), 
\eea
where $s^I(\x)$ is a boundary value and 
\bea
\nonumber
K_{\D}(w,\x)=\l \frac{w_0}{w_0^2+(\w-\x )^2} \r ^{\D}.
\eea
With this normalization of the bulk-to-boundary propagator 
the two-point function of 
corresponding boundary operators appears to be finite in the limit when 
the AdS cut-off $\eps$ tends to zero (see Appendix B for details).

Introducing the notation
\bea
D_{\D_1\D_2\D_3\D_4}=\int \frac{d^5w}{w_0^{5}}
K_{\D_1}(w,\x_1)K_{\D_2}(w,\x_2)
K_{\D_3}(w,\x_3)K_{\D_4}(w,\x_4)
\eea
and using identities for $D$-functions (see Appendix B) 
we find the following on-shell value for (\ref{ac4p}):
\bea
\la{basic}
&&S=\frac{N^2}{8\pi^2}
\int d^4x_1d^4x_2 d^4x_3d^4x_4s^{I_1}(\x_1)s^{I_2}(\x_2) s^{I_3}(\x_3)s^{I_4}(\x_4)
\biggl( \\
\nonumber
&&\frac{1}{2^7 \pi^{8}}C^-_{I_1I_2I_3I_4}
\frac{1}{x_{12}^2x_{34}^2}\l 2(x_{13}^2x_{24}^2-x_{14}^2x_{23}^2)
D_{2222}
-x_{24}^2D_{1212}-x_{13}^2D_{2121}+x_{14}^2D_{2112}
+x_{23}^2D_{1221} \r
\\
\nonumber
&&- \frac{1}{2^7 \pi^{8}}\d^{I_1I_2}\d^{I_3I_4}
\l -\frac{1}{2x_{34}^2}D_{2211}+
\frac{(x_{13}^2x_{24}^2+ x_{14}^2x_{23}^2  
 -x_{12}^2x_{34}^2)}{x_{34}^2}
D_{3322}+ \frac{3}{2}D_{2222} \r
\\
\nonumber
&&-\frac{1}{2^6 \pi^{8}}   \left.
C^{+}_{I_1I_2I_3I_4}\l \frac{1}{x_{34}^2}D_{2211}
+4x_{34}^2D_{2233}-3D_{2222} \r \right), 
\eea
where $C^{\pm}_{I_1I_2I_3I_4}=\frac{1}{2}(C_{I_1I_2I_3I_4}\pm C_{I_2I_1I_3I_4})$.
The expression under the integral represents the contribution of the 
s-channel since it possesses the s-channel symmetries 
$1\leftrightarrow 2$,  $3\leftrightarrow 4$ and $(12)\leftrightarrow (34)$.
In the expression for the 4-point function the t-channel contribution
is obtained from this one by the interchange $1\leftrightarrow 4$
and the u-channel one by $1\leftrightarrow 3$.

Taking into account the normalization of the quadratic part of 
(\ref{ac4p1}) and formula (\ref{2pt}) from the Appendix B,
we get the 2-point function of unnormalized CPOs ${\cal O}^I$:
\bea
\langle {\cal O}^{I}(\x_1){\cal O}^{J}(\x_2) \rangle
=\frac{N^2}{2^5\pi^4}\frac{\d^{IJ}}{x_{12}^4}. 
\eea
Introducing then the normalized CPOs as 
$O^I=\frac{(2^5\pi^4)^{1/2}}{N}{\cal O}^I$,
we obtain from (\ref{basic}) the following 4-point function 
of the normalized CPOs:  
\bea
\la{4p}
&&\langle O^{I_1}(\x_1)O^{I_2}(\x_2)O^{I_3}(\x_3)O^{I_4}(\x_3) \rangle
=\frac{8}{N^2\pi^2}\times \\
\nonumber
&&\biggl(-C^{-}_{I_1I_2I_3I_4}
\frac{1}{x_{12}^2x_{34}^2}\l 2(x_{13}^2x_{24}^2
-x_{14}^2x_{23}^2)D_{2222}
-x_{24}^2D_{1212}-x_{13}^2D_{2121}+x_{14}^2D_{2112}
+x_{23}^2D_{1221} \r
\\
\nonumber
&&+\d^{I_1I_2}\d^{I_3I_4}
\l -\frac{1}{2x_{34}^2}D_{2211}+
\frac{(x_{13}^2x_{24}^2+ x_{14}^2x_{23}^2  
 -x_{12}^2x_{34}^2)}{x_{34}^2}
D_{3322}+ \frac{3}{2}D_{2222} \r
\\
\nonumber
&&+2\left.
C^{+}_{I_1I_2I_3I_4}\l \frac{1}{x_{34}^2}D_{2211}
+4x_{34}^2D_{2233}-3D_{2222} \r +t+u\right), 
\eea
where $t$ and $u$ stand for the above discussed contributions of the 
$t$- and $u$-channels. Due to the conformal behaviour of the $D$-functions
Eq.(\ref{4p}) represents a correct conformally covariant expression 
for a 4-point function of operators with conformal dimension $\D=2$.

This set of 4-point functions allows one to approach the problem
of finding the OPE of the simplest CPOs in ${\cal N}=4$ SYM$_4$
that will be the subject of our further study.

\section{Appendix A} 
\setcounter{equation}{0}
{\bf Integrals of spherical harmonics} 

Considering the action for the fields $s^I$,
we need the following explicit expressions for the integrals $a_{I_1I_2I_3}$ 
and $t_{I_1I_2I_3}$ involving the scalar
spherical harmonics $Y^I$ \footnote{They describe a basis of 
irrep. {\bf 20} of $SO(6)$.} 
and Killing vectors $Y^I_a$ \cite{LMRS,AF5}:
\bea
\la{int}
a_{I_1I_2I_3}=\frac{2^2\cdot 6^{1/2}}{5\pi^{3/2}}C_{I_1I_2I_3}~~~~
t_{I_1I_2I_3}=\frac{6^{1/2}}{\pi^{3/2}}T_{I_1I_2I_3}.
\eea
If we introduce a basis $C_{ij}^I$ in the space of symmetric traceless 
second rank tensors of $SO(6)$ and a basis 
$C_{i;j}^I$ in the space of antisymmetric tensors 
with normalization conditions 
\bea
\la{norm}
C_{ij}^IC_{ij}^J=\d^{IJ},~~~~C_{i;k}^IC_{j;k}^J=\frac{1}{6}\d^{IJ}\d_{ij}
\eea
then the tensors $C_{I_1I_2I_3}$ and  $T_{I_1I_2I_3}$
are given by 
\bea
\nonumber
C^{I_1I_2I_3}=C^{I_1}_{ij}C^{I_2}_{jk}C^{I_3}_{ki},~~~
T^{I_1I_2I_3}=C^{I_1}_{ik}C^{I_2}_{kj}C_{i;j}^{I_3}
-C^{I_1}_{jk}C^{I_2}_{ki}C_{i;j}^{I_3},
\eea
where we have written tensor $T^{I_1I_2I_3}$ to be 
explicitly antisymmetric 
in indices $I_1,I_2$.

One can easily establish the following summation formula 
\bea
\la{sum1}
\sum_{I}C^I_{ij}C^I_{kl}=\frac{1}{2}\d_{ik}\d_{jl}
+\frac{1}{2}\d_{il}\d_{jk}-\frac{1}{6}\d_{ij}\d_{kl}
\eea
that steams from the fact that the l.h.s. of the expression 
above is a fourth rank tensor 
of $SO(6)$, symmetric and traceless both in $(ij)$ and $(kl)$ 
indices with the normalization
condition $C^{I}_{ij}C_{ij}^I=20$.

Analogously one finds 
\bea
\la{sum2}
\sum_{I}C_{m;l}^{I}C_{n;s}^{I}
=\frac{1}{2}(\d_{mn}\d_{ls}-\d_{ms}\d_{nl})
\eea
since this time the l.h.s. of (\ref{sum2}) is a 
traceless and antisymmetric
in $m,l$ and in $n,s$ indices fourth rank tensor of 
$SO(6)$ that agrees with the normalization (\ref{norm}). 

\vskip 0.3cm
\noindent {\bf Some properties of $SO(6)$ $\G$-matrices}

In studying the action of the gauged supergravity, 
we need an identity that 
follows from the completeness condition for $SO(6)$ $\G$-matrices
and may be found in \cite{GRW1,PPN,GRW}.
To make the treatment self-contained we recall its derivation here.

Consider the Clifford algebra in $d=6$ Euclidean dimensions:
$$
\{\G_i,\G_j\}=2\d_{ij},~~~~i,j,k,l,n=1,...,6.
$$
The $\G$-matrices can be represented by hermitian 
skew-symmetric $8\times 8$ matrices 
$(\G_i)_a^b$. Indices $a,b=1,...,8$ are raised or lowered 
by the symmetric charge conjugation matrix 
$C_{ab}$ that in the chosen representation coincides with $\d_{ab}$. 
Thus, we do not distinguish the upper and lower indices.

Clearly, the matrices 
\bea
\la{b}
\G_i,~~~i\G_i\G_0,~~~\G_{ij},~~~\G_0=i\G_1\G_2\G_3\G_4\G_5\G_6
\eea
are skew-symmetric. Their number is $6+6+15+1=28$ and 
it coincides with 
a total number $8\cdot 7/2=28$ of independent skew-symmetric 
matrices among all $8\times 8$
matrices. Therefore, any skew-symmetric matrix $A_{ab}$ 
can be decomposed over the basis (\ref{b}):
\bea
A_{ab}=\a_1^{i}(\G_i)_{ab}+\a_2^{i}(i\G_i\G_0)_{ab}
+\frac{1}{2}\a_3^{ij}(\G_{ij})_{ab}+\a_4(\G_0)_{ab}.
\la{anm}
\eea
Here in the third term we assume the summation over the 
whole set of indices - not just over 
$i<j$. We also use the convention that $\a_3^{ij}=-\a_3^{ji}$. 
The coefficients 
are easy to compute 
$$
\a_1^{i}=\frac{1}{8}tr(A\G_i),~~~
\a_2^{i}=\frac{i}{8}tr(A\G_i\G_0),~~~
\a_3^{ij}=-\frac{1}{8}tr(A\G_{ij}),~~~
\a_4=\frac{1}{8}tr(A\G_0).
$$
Substituting these coefficients back in (\ref{anm}), 
and using the fact that 
Eq.(\ref{anm}) should hold for any skew-symmetric 
matrix $A_{ab}$, we find an identity:
\bea
\nonumber
\frac{1}{16}(\G_{ij})_{ab} (\G_{ij})_{cd}
-\frac{1}{8}(\G_i)_{ab}(\G_i)_{cd}
-\frac{1}{8}(i\G_i\G_0)_{ab}(i\G_i\G_0)_{cd}
=\frac{1}{2}(\d_{ac}\d_{bd}-\d_{ad}\d_{bc})-
\frac{1}{8}(i\G_0)_{ab}(i\G_0)_{cd},
\eea
the term with $\a_4$ was written in the l.h.s..

If one introduces the symplectic metric $\O^{ab}=-i(\G_0)^{ab}=-\O_{ab}$ 
and matrices $\G_{i\a}=(\G_{i},i\G_i\G_0)$ for $\a=1,2$ then
the last indentity reads as follows \cite{GRW1,PPN,GRW}:
\bea
\la{id}
\frac{1}{16}(\G_{ij})_{ab} (\G_{ij})^{cd}
-\frac{1}{8}(\G_{i\a})_{ab}(\G_{i\a})^{cd}
=\frac{1}{2}(\d_a^c\d_b^d-\d_a^d\d_b^c)
+\frac{1}{8}\O_{ab}\O^{cd}.
\eea
Here in the l.h.s. we have written some indices up
since the r.h.s. represents now a tensor of $USp(8)$. 
It is as well to note that except the symmetric charge 
conjugation matrix that is just the unit matrix one can also  
raise and lower indices with the $USp(8)$ metric $\O_{ab}$.

We also summarize the trace formulae needed in the paper 
\bea
&&\tr(\G_{ij}\G_{kl})=8(\d_{il}\d_{kj}-\d_{ik}\d_{jl}), \la{tr1} \\
&&\tr(\G_{in}\G_{jn}\G_{kl})=32(\d_{ik}\d_{jl}-\d_{il}\d_{jk}), \la{tr2} \\
&&\tr(\G_{i\a}\G_{j\a}\G_{kl})=16(\d_{il}\d_{jk}-\d_{ik}\d_{jl}). \la{tr3}
\eea 

Note that matrices $\G_i$ are hermitian while 
$\G_0$, $\G_{ij}$ and $i\G_i\G_0$ are antihermitian.
It follows from here that $\G_{ij}$ and $i\G_i\G_0$ are real. 

\vskip 0.3cm
\noindent {\bf Scalar kinetic part of the 
lagrangian of the gauged 5d supergravity}

By using (\ref{id}), one can check the following relation:
\bea
(V^{-1})_{cd}^{~AB}V_{AB}^{~ab}=(V^{-1})_{cd~ij}V^{ij~ab}
+(V^{-1})_{cd}^{~i\a}V_{i\a}^{~ab}=
\frac{1}{2}(\d_a^c\d_b^d-\d_a^d\d_b^c)+\frac{1}{8}\O_{ab}\O^{cd}
\eea
that is an $USp(8)$ analog of $VV^{-1}=I$. 
The properties of the $\G$-matrices, in particular, (\ref{tr1})
imply the further relations:
$$
(V^{-1})_{ab~kl}V^{ij~ab}=
\frac{1}{2}(\d_k^i\d_l^j-\d_k^j\d_l^i),~~~
(V^{-1})_{ab}^{~i\a}V_{j\b}^{~ab}=\d_i^j\d_{\a}^{\b}
$$
and also 
$$
(V^{-1})_{ab}^{~i\a}V_{kl}^{~ab}=(V^{-1})_{ab~kl}V^{~i\a~ab}=0.
$$
In the $SL(6,{\bf R})\times SL(2,{\bf R})$ basis the 
element $P_{\mu~ab}^{~~~~cd}$ is given by 
\bea
\nonumber
P_{\mu~ab}^{~~~~cd}&=&(V^{-1})^{cd}_{~ij}\n_{\mu}V^{ij}_{~ab}
+(V^{-1})^{cd~i\a}\n_{\mu}V_{i\a~ab} \\
&+&2Q_{\mu [a}^{[c}\d_{b]}^{d]}
+gA_{\mu i}^{~~j}\l 2V^{ik}_{~ab}(V^{-1})_{jk}^{~cd}-(V^{-1})^{cd~i\a}V_{j\a~ab}\r .
\eea
If we now require that $P_{\mu~ab}^{~~~~cd}$ is in 
the coset space $E_6/USp(8)$, then the trace 
$P_{\mu~ab}^{~~~~cb}$ should be equal to zero. 
This allows one to solve 
$Q_{\mu [a}^{[c}\d_{b]}^{d]}$ via the vielbein:
\bea
Q_{\mu a}^{~~b}=-\frac{1}{3}\l (V^{-1})^{bc~AB}\n_{\mu}V_{AB~ac}+
g A_{\mu i}^{~~j}(2V^{ik}_{~ac}(V^{-1})_{jk}^{~~bc}
-V_{j\a~ac}(V^{-1})^{bc~i\a} \r
\eea
Substitution of the explicit expressions (\ref{viel}) yields
\bea
\la{con}
Q_{\mu a}^{~~~b}&=&\frac{1}{24}\l \G_{in}\G_{jn}-\G_{i\a}\G_{j\a} \r _a^{~b}
(\n_{\mu}SS^{-1}+g S A_{\mu}S^{-1} )_i^{~j},
\eea 
where on the r.h.s. the expression for the matrix 
$R_{\mu}$ defined by (\ref{R}) appeared.

It is useful to note the following summation formula for $\G$-matrices:
$$
\G_{in}\G_{jn}-\G_{i\a}\G_{j\a}=-6\G_{ij}-7\d_{ij}\cdot I.
$$
Upon substituting this in (\ref{con}), the term with $\d_{ij}$
vanishes due to the tracelessness of $R_{\mu }$.
Thus, we finally get 
\bea
\la{conf}
Q_{\mu a}^{~~~b}&=&\frac{1}{4}(\G_{ij})_a^{~b}R_{\mu i}^{~~j}.
\eea
It is easy to see that $Q_{\mu a}^{~~~b}$ is an antihermitian 
matrix indeed being an element 
of $Usp(8)$ Lie algebra, i.e., obeying the condition
$$
Q_{\mu a}^{~~~b}=-\O^{bc}Q_{\mu c}^{~~~d}\O_{da}.
$$
For the element $P_{\mu ab}^{~~~cd}$ we, therefore, get 
\bea
\la{tenP}
P_{\mu ab}^{~~~cd}=\frac{1}{8}
\l (\G_{in})^{cd}(\G_{jn})_{ab}-(\G_{i\a})^{cd}(\G_{j\a})_{ab}\r 
R_{\mu i}^{~~j}+
2Q_{\mu [a}^{~~~[c}\d_{b]}^{d]}.
\eea

Since tensor $P_{\mu ab}^{~~~cd}$ is completely fixed by the condition of 
the vanishing trace one now can check that (\ref{tenP})  
is indeed an element orthogonal to 
$USp(8)$-part of the Lie algebra of $E_{6(6)}$ w.r.t. to the Killing metric.
Orthogonality means the following relation 
\bea
\la{ort}
P_{\mu ab}^{~~~cd}U_{cd}^{~~~ab}=0,
\eea 
where $U_{cd}^{~~~ab}=Q_{\mu [c}^{~~~[a}\d_{d]}^{b]}$ is an element of the 
$USp(8)$ Lie algebra. Formula (\ref{ort}) then easily follows from 
(\ref{tr1}), (\ref{tr2}) and the relation
\bea
\nonumber
&&2Q_{\mu [a}^{~~~[c}\d_{b]}^{d]}Q_{\mu [c}^{~~~[a}\d_{d]}^{b]}=
3Q_{\mu a}^{~~~c}Q_{\mu c}^{~~~a}=
-\frac{3}{2}(R_{\mu i}^{~~j}R_{\mu i}^{~~j}-R_{\mu i}^{~~j}R_{\mu j}^{~~i}).
\eea

We  also need
$(P_{\mu})^{ab}_{~~~cd}=
\O^{aa'}\O^{bb'}\O_{cc'}\O_{dd'}P_{\mu a'b'}^{~~~c'd'}$:
\bea
\nonumber
(P_{\mu})^{ab}_{~~~cd}=
\frac{1}{8}\l (\G_{in})_{cd}(\G_{jn})^{ab}-(\G_{i\a})_{cd}(\G_{j\a})^{ab}\r 
R_{\mu j}^{~~i}-2Q_{\mu [c}^{~~~[a}\d_{d]}^{b]}.
\eea
Now we are ready to compute the scalar kinetic part of 
Lagrangian (\ref{ags}).
By using the orthogonality condition (\ref{ort}) 
we can write it in the form
\bea
\nonumber
P_{\mu abcd}P_{\mu}^{abcd}&=&\l
\frac{1}{8}\l (\G_{in})^{cd}(\G_{jn})_{ab}
-(\G_{i\a})^{cd}(\G_{j\a})_{ab}\r 
R_{\mu j}^{~~i}+2Q_{\mu [a}^{~~~[c}\d_{b]}^{d]}\r \\
\nonumber
&\times &
\l \frac{1}{8}\l (\G_{km})_{cd}(\G_{lm})^{ab}
-(\G_{k\b})_{cd}(\G_{l\b})^{ab}\r 
R_{\mu l}^{~~k} \r.
\eea
After some algebra we arrive at the answer  
\bea
P_{\mu abcd}P_{\mu}^{abcd}&
=&3R_{\mu i}^{~~j}(R^{\mu})_i^{~~j}
+3R_{\mu i}^{~~j}
(R^{\mu})_j^{~~i}=\frac{3}{2} \tr{\l R_{\mu}+ R_{\mu}^t \r^2 }.
\eea
Note that the r.h.s. of the scalar kinetic term appears to be manifestly 
positive in an Euclidean signature space as it should be.

\section{Appendix B} 
\setcounter{equation}{0}
\noindent {\bf z-integrals}

$z$-integrals are computed by using the technique by \cite{HFR}. 
We list here the corresponding results:
\bea
\nonumber
&&\int \frac{d^{5}z}{z_0^5}G_{\D}(u)s^{I_3}(z)s^{I_4}(z)=
\frac{1}{2^4\pi^4}
\int d^4x_3d^4x_4\frac{s^{I_3}(\x_3)s^{I_4}(\x_4)}
{x_{34}^2}K_1(w,\x_3)K_1(w,\x_4), \\
\nonumber
&&\int \frac{d^5 z}{z_0^5}
G_{\mu\nu}(u)s^{I_3} \stackrel{\leftrightarrow~}{\n^\nu}s^{I_4}(z)=
\frac{1}{2^4 \pi^4}\int d^4x_3d^4x_4
\frac{s^{I_3}(\x_3)s^{I_4}(\x_4)}{x_{34}^2}
K_1(w,\x_3)\stackrel{\leftrightarrow~}
{\n_{\mu}} K_1(w,\x_4), \\
\nonumber
&&\int \frac{d^5z}{z_0^5} G_{\mu\nu~\rho\lambda}(u)T^{\rho\lambda}(z)=\frac{1}{2^4\pi^4} 
\int d^4x_3d^4x_4 \frac{s^{I}(\x_3)s^{I}(\x_4)}{x_{34}^2} \\
&&\nonumber
\l g_{\mu\mu'}g_{\nu\nu'}+g_{\mu\nu'}g_{\nu\mu'}
-\frac{2}{3}g_{\mu\nu}g_{\mu'\nu'}\r  
\n^{\mu'} K_1(w,\x_3) \n^{\nu'} K_1(w,\x_4).
\eea
Note that computing the last integral, 
we have used the gauge freedom
in the definition of the graviton propagator 
to obtain the answer in the  
simplest covariant form.

\vskip 0.3cm
\noindent {\bf Two-point function of lowest weight CPOs}

As was noted in \cite{FMMR}, a correct way to compute 
a two-point correlation function of 
operators in the boundary CFT, which is compatible 
with the Ward identitites,
consists of two steps. First one uses the prescription 
by \cite{GKP} for posing 
the Dirichlet boundary problem on gravity fields. 
Then one computes the two-point function 
in the momentum space and transform it further to the $x$-space.
Below we undertake this procedure to find the 
two-point function of the lowest weight CPOs.
 
For a scalar field of the AdS-mass $m^2=-4$ with the conventionally
normalized quadratic action,   
the solution of the Dirichlet boundary problem reads as 
\bea
\nonumber
K(z,k)=\l \frac{z_0}{\eps} \r^{2}\frac{K_0(kz_0)}{K_0(k\eps)}
\eea 
with the Fourier transform defining the following bulk-to-boundary propagator
\bea
\la{prop}
K(z,\x)=-\frac{1}{2\pi^2\eps^2\ln{\eps}}\l \frac{z_0}{z_0^2+|\x|^2}  \r^2
=-\frac{1}{2\pi^2\eps^2\ln{\eps}}K_2(z,\x).
\eea
For the two-point correlation function in the 
momentum space we then have \cite{FMMR}:
\bea
\nonumber
&&\langle O(\vk )O(\vk ')\rangle=\eps^{-3}\d(\vk +\vk ')\lim_{z_0\to \eps}
\p_{z_0}\l \l \frac{z_0}{\eps} \r^{2}\frac{K_0(kz_0)}{K_0(k\eps)}
\r =\d(\vk +\vk ')\frac{k}{\eps^3}\frac{K_1(k\eps)}{K_0(k\eps)}.
\eea
where a nonessential local term $1/\eps^4$ was 
omitted and $k$ denotes $|\vec{k}|$.
Decomposing the result in power series, one gets  
\bea
\nonumber
\langle O(\vk )O(\vk ')\rangle &=&-\d(\vk +\vk ')\frac{k}{\eps^3}
\frac{\frac{1}{k\eps}
+\sum_{n=0}^{\infty}\frac{(k\eps/2)^{2n+1}}{n!(n+1)!}\l \ln\frac{k\eps}{2}
-\frac{1}{2}\psi(k+1)-\frac{1}{2}\psi(k+2)  \r}
{-\ln\eps-\ln k +\ln 2-\psi(k+1)+\eps(...)  }\\ 
\nonumber
&&=\d(\vk +\vk ')\frac{1}{\eps^4\ln{\eps}}\l 1-\frac{\ln k}{\ln \eps}+ \frac{k^2\eps^2}{2}\ln{k}+... \r .
\eea
The most singular relevant term here is the second one, so modulo 
local terms one finds 
\bea
\nonumber
\langle O(\vk )O(\vk ')\rangle &=&-\d(\vk +\vk ')
\frac{1}{\eps^4\ln^2\eps } \ln k .
\eea
Performing the Fourier transform, we finally get  
\bea
\la{one}
\langle O(\x_1)O(\x_2)\rangle &=&\frac{1}{2\pi^2\eps^4\ln^2\eps x_{12}^4}.
\eea
In order to have a finite 2-point function in the limit 
$\eps\to 0$ one has to rescale the boundary operator as $O(\x)\to -\frac{1}{\eps^2\ln{\eps}}O(\x)$,
so that 
\bea
\la{2pt}
\langle  O(\x_1) O (\x_2)\rangle &=&\frac{1}{2\pi^2 x_{12}^4}.
\eea
To preserve the scale-invariance of the interaction 
term $\int d^4x O(\x)s(\x)$, where
$s(\x)$ is the boundary value of the bulk supergravity 
scalar $s(z)$ we then need to rescale 
the $s(\x)$ in a way $s(\x)\to -\eps^2\ln{\eps}s(\x)$. After 
this rescaling the solution
of the Dirichlet boundary problem  reads as (\ref{sDp}).

\vskip 0.3cm
\noindent {\bf Some identitites for $D$ functions}

As soon as $z$-integrals are performed, one is left with contact diagrams 
involving different numbers of derivatives. By using the identity \cite{HF1}  
\bea
\nonumber
\n_{\mu}K_{\D_1}(w,\x_1)\n^{\mu}K_{\D_2}(w,\x_2)=
\D_1\D_2\l K_{\D_1}(w,\x_1)K_{\D_2}(w,\x_2)
-2 x_{12}^2K_{\D_1+1}(w,\x_1)K_{\D_2+1}(w,\x_2) \r
\eea
all the contact diagrams are then reduced to the sum of different 
$D$-functions.

In \cite{HFMMR} some identities involving different $D$-functions were proved.
We made use of the following ones:
\bea
\nonumber
&&x_{24}^2D_{2312}+x_{23}^2D_{2321}=D_{2211}-2x_{12}^2D_{3311}, \\
\nonumber
&&2x_{12}^2D_{3311}=\frac{1}{2}x^2_{34}D_{2222}+\frac{1}{2}D_{2211}, \\
\nonumber
&&x_{24}^2D_{1212}=x_{13}^2D_{2121},~~~x_{14}^2D_{2112}=x_{23}^2D_{1221}\\
\nonumber
&&x_{13}^2x_{12}^2D_{3221}+x_{24}^2x_{34}^2D_{1223}=-
\frac{1}{2}(x_{12}^2x_{34}^2+x_{13}^2x_{23}^2)D_{2222} \\
\nonumber
&&-\frac{3}{2}x_{14}^2D_{2112}+2x_{14}^4D_{3113}+\frac{1}{2}B.
\eea
and identities obtained from these by different permutations of indices
to reduce the number of possible $D$-functions appearing
in the 4-point function of the lowest weight CPOs to the minimal set 
giving by $D_{1212}$, $D_{2233}$ (with different permutations of indices)
and $D_{2222}$. Here $B$ is a generating function for $D_{\D_1\D_2\D_3\D_4}$
and it is given by 
$$
B=\frac{\pi^2}{2}\int \frac{\prod d\alpha_j\d(\sum \alpha_j-1)}
{(\sum \alpha_k\alpha_l x_{kl}^2)^2}.
$$

\vskip 1cm
{\bf ACKNOWLEDGMENT}

G.A. is grateful to S. Theisen, S. Kuzenko and to A. Petkou, 
and S.F. is grateful
to A. Tseytlin and S. Mathur
for valuable discussions.
The work of G.A. was
supported by the Alexander von Humboldt Foundation and in part by the
RFBI grant N99-01-00166, and the work of S.F. was supported by
the U.S. Department of Energy under grant No. DE-FG02-96ER40967 and
in part by RFBI grant N99-01-00190.

\end{document}